# Gunn Effect in Silicon Nanowires: Charge Transport under High Electric Field

Daryoush Shiri[1], Amit Verma[2], Reza Nekovei[2], Andreas Isacsson[1], C. R. Selvakumar[3], and M. P. Anantram[4]

[1]Department of Physics, Chalmers University of Technology, SE-41296 Göteborg, Sweden

[2]Department of Electrical Engineering and Computer Science, Texas A&M University-Kingsville, Kingsville, Texas 78363, USA

[3]Department of Electrical and Computer Engineering, University of Waterloo, Waterloo, Ontario N2L 3G1, Canada

[4]Department of Electrical Engineering, University of Washington, Seattle, Washington 98195-2500, USA

**Corresponding Author's Email:** shiri@chalmers.se

**Abstract:** Gunn (or Gunn-Hilsum) Effect and its associated negative differential resistivity (NDR) emanates from transfer of electrons between two different energy bands in a semiconductor. If applying a voltage (electric field) transfers electrons from an energy sub band of a low effective mass to a second one with higher effective mass, then the current drops. This manifests itself as a negative slope or NDR in the I-V characteristics of the device which is in essence due to the reduction of electron mobility. Recalling that mobility is inversely proportional to electron effective mass or curvature of the energy sub band. This effect was observed in semiconductors like GaAs which has direct bandgap of very low effective mass and its second indirect sub band is about 300 meV above the former. More importantly a self-repeating oscillation of spatially accumulated charge carriers along the transport direction occurs which is the artifact of NDR, a process which is called Gunn oscillation and was observed by J. B. Gunn. In sharp contrast to GaAs, bulk silicon has a very high energy spacing (~1 eV) which renders the initiation of transfer-induced NDR unobservable. Using Density Functional Theory (DFT), semi-empirical 10 orbital ($sp^3d^5s^*$) Tight Binding (TB) method and Ensemble Monte Carlo (EMC) simulations we show for the first time that (a) Gunn Effect can be induced in narrow silicon nanowires with diameters of 3.1 nm under 3 % tensile strain and an electric field of 5000 V/cm, (b) the onset of NDR in I-V characteristics is reversibly adjustable by strain and (c) strain can modulate the value of resistivity by a factor 2.3 for SiNWs of normal I-V characteristics i.e. those without NDR. These observations are promising for applications of SiNWs in electromechanical sensors and adjustable microwave oscillators. Results are explained using electron-phonon scattering mechanisms including both Longitudinal Acoustic (LA) and Optical (LO) phonons involving both inter-sub band and intra-sub band scattering events. It is noteworthy that the observed NDC is different in principle from Esaki-Diode and Resonant Tunneling Diode (RTD) type of negative I-V slopes which originate from tunneling effect.

**KEYWORDS**: Gunn Diode, Negative Differential Resistivity (NDR), Strained silicon nanowires, Electron-phonon scattering, Microwave oscillator, Charge carrier mobility, Hot electron.

**INTRODUCTION**

Silicon is predicted to remain as the fundamental material for the micro- and nano-electronics industry and ultra large scale integrated (ULSI) circuits [1] despite the ever growing research devoted to other materials and their nano structured variants. Silicon Nanowires (SiNW) are one of the members of silicon nanostructure family, which have so far shown promising applications in areas ranging from biological sensors [2], thermoelectric converters [3], opto-mechanical devices [4], piezoelectric sensors [5], and solar cells [6] among others. Ease of bandgap conversion from indirect to direct one with crystallography [7], mechanical strain [8-11], and alloying bring SiNWs into realm of optical applications e.g. photodetectors [12-15] and light emitters [16]. In addition, breaking the centro-symmetricity in strained nanowires was shown to enhance the second order nonlinear optical susceptibility suitable for second harmonic or frequency difference (THz) generation applications [17-20]. With the advent of new top-down CMOS compatible fabrication methods for silicon nanowires [21, 22] it is now feasible to build spin-based quantum gates [23], spintronic devices [24-26], MOSFETs with critical dimensions approaching 5 nm or less [27-29], logic circuits [30], memory [31] and memristive devices [32, 33]. Readers are referred to references [34, 35] for an extensive review of the field. To add more to the list of SiNW applications and functionalities, looking back at differences of bulk silicon and their III-V counterparts is instructive. We have to ask what are the other properties which bulk silicon could not possess and yet they might be achievable in silicon nanowire due to its peculiar band structure and its adjustability?

Our answer to this question was "*Gunn (or Gunn-Hilsum) Effect*" and in this article we show for the first time that this effect and its associated negative differential resistivity (NDR) is observable in silicon nanowires. The negative differential resistivity which is reduction of current density (*j*) with an increase in the electric field (*E*), arises when the scattering rate of electrons by lattice vibrations (phonons) or impurities is enhanced by electric field. Recall that J and E are related as j = σ.E wherein σ is the conductivity tensor. Among these scattering processes the *electron transfer mechanism* is the one which involves migration of electrons from a low effective mass sub band to another sub band of higher effective mass which is mediated by the momentum of phonons as the field increases. The difference between the energies of these sub bands (Δ$E$) must be reasonably high e.g. a few $k_B T$ to ensure that this upper valley or sub band is yet unpopulated at room temperature for low electric field values. Here $K_B$ and *T*, are Boltzmann constant and the absolute temperature in Kelvin, respectively. On the other hand, *ΔE* has to be small enough to avoid dielectric breakdown under an intense electric field. Figure 1.a shows the band structure of bulk GaAs in which the valley splitting or offset between the energies of *Γ* and L valleys is about 300 meV. The *Γ* valley has a very low effective mass (0.0632$m_0$) where $m_0$ is the free mass of electron. For low to moderate electric fields, most electrons are within this sub band (valley) and contribute to large electron mobility ($\mu_H \sim$ 8000 cm$^2$/V.s) [36]. As populating the L valley with high effective mass (0.55$m_0$) occurs for higher electric fields (> 3000 V/cm) along the <111> crystallographic direction, the mobility, and as a result the current, reduce ($\mu_L \sim$ 2500 cm$^2$/V.s) [See Figure 1.b]. This leads to an I-V characteristics or velocity versus electric field as shown in Figure 1.c which has a NDR section. This mechanism of obtaining NDR in general was proposed by Ridley [37] and Hlisum [38] and later as Ehrenreich experimentally obtained the band structure of GaAs, the search for materials of suitable band structure for this type of NDR has begun [39]. As the negative differential resistivity of the GaAs (or as it was later called Gunn diode) can compensate the Ohmic loss in electric circuits, it was proposed to be applied in electric oscillators of microwave frequencies (Figure 1.c).

However, the aforementioned electron transfer mechanism is lacking in bulk silicon, the reason of which is evident by looking at its band structure in Figure 1.d. As it can be seen, the offset between X and L sub bands is about 1 eV which mandates a very high voltage to induce NDR. Uniaxial straining of p-type doped silicon was suggested to reduce this energy offset and initiate NDR [37]. Unaware of the theoretical and experimental investigations of high electric field effect on charge transport in GaAs, J. B. Gunn observed instabilities in the current flowing through a piece of GaAs under high voltage[40], the source of which was the same I-V characteristics of Figure 1.c. He observed that when the electric field reaches to a threshold value of 2000-4000 V/cm, the current (drift velocity) of electron drops and an accumulation layer is built up after which there is a depleted region (positively charged) since the electrons left that section with their high speed [See Figure 1.e].

The internal dipolar field of such accumulation/depletion sandwich reduces the total external field and reverts the current back to its original value. After the dipolar region reaches to one end of the device, i.e. moves from cathode and arrives at anode, the voltage across the device increases again, hence the current increases to the point at which the same NDR drop occurs and the procedure repeats itself. This results in an oscillating current around an average DC value [Figure 1.e]. This self-repeating increase and decrease of current can reach up to 12 GHz in frequency depending on the length and doping density. Length and drift velocity of electrons determine the transit time through the sample or the frequency of current oscillation. Oscillations of the same origin were observed in materials like CdTe, GaAs$_{1-x}$P$_x$, InP and GaSb [39]. In addition to exploiting this effect as an intrinsic source of microwave oscillations, the resulting NDR in I-V characteristics, can also be used to compensate the Ohmic loss of an electrical resonator (Figure 1.c). The oscillator of this type is called Gunn diode oscillator and is used in microwave applications with frequencies of up to 100 GHz depending on the required output power [41].

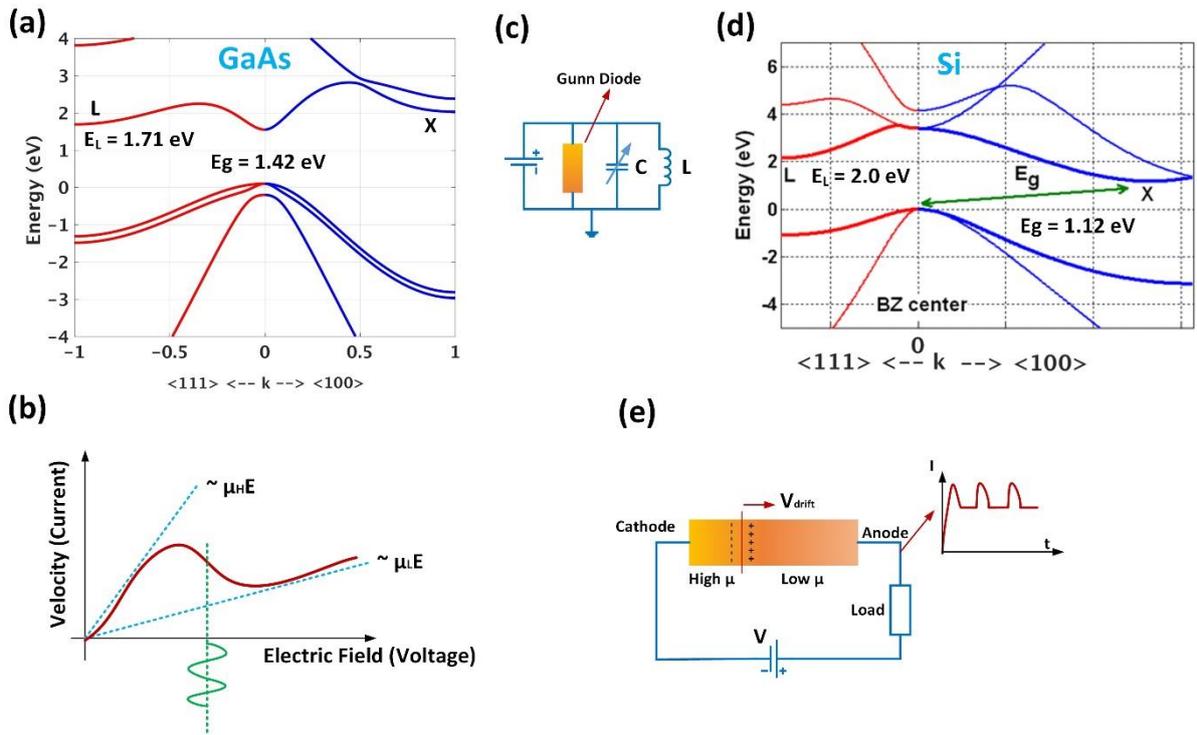

**Figure 1. Comparing the band structure of bulk GaAs and Si and NDR-induced oscillation in a Gunn diode.** Band structures of (a) Gallium Arsenide with direct bandgap. Direct and indirect sub bands are called Γ and L valley, respectively. The energy offset is 300 meV. (b) Negative Differential Resistance (NDR) for GaAs as a result of velocity drop for electrons which migrated to high effective mass (L) band. The NDR emanated from a large ratio of effective mass of indirect to direct conduction sub band which is about 100 for GaAs. (c) The electric network shows how compensating the loss of an electric LC resonator using a Gunn diode with NDR results in a perpetual oscillation in microwave frequencies. (d) Band structure of bulk silicon with more than 1 eV offset between Γ and L valleys which explains why there is no NDR in bulk silicon. (e) The second type of Gunn oscillation (intrinsic) is a result of self-repeating formation of accumulation/depletion sandwich inside the bulk material moving with saturated drift velocity from cathode to anode.

Motivated by previous theoretical and experimental observations of indirect to direct bandgap conversion in SiNWs due to mechanical strain [8][42-43] and the advent of novel nanofabrication methods to put these devices into work [5][21][44], we investigated the possibility of observing Gunn-type NDR in SiNWs. In contrast to bulk silicon in which no Gunn Effect was observed under high electric field (below breakdown field), in this article for the first time we show that; (a) Gunn Effect or NDR emerges in narrow [110] SiNWs under tensile strain, (b) The threshold voltage (field) for this NDR emergence is tunable by the strain, i.e., NDR can be inhibited or initiated reversibly, and (c) the conductivity of the SiNWs before NDR occurrence is also tunable by mechanical strain as a result of direct to indirect bandgap conversion and corresponding effective mass change. This leads to a factor of 2.3 change in the resistivity as the SiNW is strained between 0 % and -3 %. The NDR observed here is different in essence from the NDR which arises in Esaki [45, 46] as well as Resonant Tunneling Diodes (RTDs) [47, 48]. The abovementioned observations, i.e., (a), (b) and (c), promise application of SiNWs in tunable microwave oscillators and mechanical (force, pressure) sensors. The rest of this article is organized by explaining the computational methods including energy minimization, band structure, and electron-phonon scattering rate calculations, followed by Ensemble Monte Carlo (EMC) simulations. Thereafter, we will discuss the observed results and prospects of an experimental proof-of-concept device.

**METHODS**

**Energy Minimization and Band structure:** Relaxation of the structural energy of SiNWs as well as mechanical strain application are performed using Density Functional Theory (DFT) method as implemented in SIESTA® [49]. The hydrogen terminated nanowires have [110] crystallographic direction due to their proven relative stability compared to [100] direction [50]. To avoid the inherent bandgap underestimation in DFT and diameter sensitive many–body GW corrections, the band structure and eigenstates were calculated with 10 orbital *(sp$^3$d$^5$s$^*$)* Tight Binding (TB) method using the parameters given in reference [51]. The TB method has shown success in reproducing the experimental data of the effect of radial strain on the photoluminescence spectra in narrow SiNWs [52, 53]. The resulting unit cell after each minimization step is fed to the next minimization round by increasing or decreasing the unit cell length (*a*) according to the desired strain amount (ε), i.e., *± aε* as shown in Figure 2.a. During these steps the volume of the unit cell is kept fixed in order to let the desired strain take effect. A detailed EMC simulator was developed that incorporated the four lowest conduction sub bands of the calculated band structure of SiNWs (strained & unstrained), to study the effect of electron-phonon scattering under high electric fields.

**Scattering Rates and Ensemble Monte Carlo (EMC) simulations:** The electron-phonon scattering rates were calculated with first order perturbation theory using Fermi's Golden rule [54]. The total scattering rate for a given state within a Brillouin Zone (BZ) with a wave vector of $k_z$, is written as a sum over all individual scattering rates from this state to available states $k_{z'}$:

$$W_{k_z} = \sum_{k_z'} W(k_z, k_z'). \qquad (1)$$

Here, W($k_z$, $k_{z'}$) is the scattering rate of an electron from its initial state at $k_z$ to a possible final (secondary) state (at $k_{z'}$). W($k_z$, $k_{z'}$) includes both inter- and intra-sub band electron-phonon scattering events. The secondary (final) states corresponding to each initial state are determined by the phonon type. For Longitudinal Acoustic (LA) phonons, all final states lie inside a window of Ez ± E$_{Debye}$ (starting from initial state energy, E$_z$), and constitute all possible k$_{z'}$. This is graphically shown in Figure 2.b for an initial state in sub band 1 (B$_1$), with all possible secondary states in sub band 3 (B$_3$) marked by circles. E$_{Debye}$ = 63 meV is the Debye energy or the maximum energy of LA phonons, and within the Debye energy window (shaded strip on Figure 2.b), the secondary states form a quasi-continuum. In practice however, the number of secondary states is limited by the resolution with which the BZ is discretized along k$_z$. For each individual term of equation (1) due to electron-LA phonon scattering, the corresponding rate, i.e., W (k$_z$, k$_{z'}$), is written as follows:

$$W(k_z, k'_z) = \frac{D_e^2}{8\pi^2 \rho \hbar^3 v_s^4} \Delta E_{kk'}^2 B_\pm \left( \left| \pm \frac{\Delta E_{kk'}}{\hbar v_s} \right| \right) \Phi(q_t, q_z). \qquad (2)$$

Here D$_e$, ρ and v$_s$ are electron deformation potential (D$_e$= 9.5 eV), mass density (ρ= 2329 Kg/m$^3$) and velocity of sound in silicon (v$_s$=9.01×10$^5$ cm/s), respectively. $B_\pm(|\tilde{q}|)$ is the Bose-Einstein factor of phonons and it is $1/(e^{\frac{\hbar \omega_q}{K_B T}} - 1)$ for absorption and $1 + \frac{1}{(e^{\frac{\hbar \omega_q}{K_B T}}-1)}$ for emission of a phonon, where $\hbar \omega_q$ is the energy of a phonon with a wave vector of **q**. Φ is a structural factor involving inter-atomic and intra-atomic matrix elements of terms like $e^{-i\tilde{q}.(r_{m'}-r_m)}$, where *r* is the coordinate of an atom. The indices m and m' are indices of different atoms within a unit cell. Details of deriving Φ is explained in the Supplementary Information and reference [55]. ΔE$_{kk'}$ is the energy difference between initial and final states, and q$_t$ and q$_z$ are transverse and longitudinal wave vectors of phonons, respectively. When electron scattering emanates from a Longitudinal Optical (LO) phonon, the secondary states are those which lie exactly above or below the energy of the initial state (at k$_z$) by E$_{LO}$. Figure 2.c depicts an example of finding available secondary states for an electron in B$_1$ which is scattered to all other sub bands including itself (intra-band scattering) assisted by LO phonons. As the optical phonon branch is almost flat and independent of wave vector, then it is possible to assume a dispersion less or flat (i.e., k$_z$ independent) line on which all LO phonons have the same energy of E$_{LO}$ = 54 meV. The individual rate in equation (1), *W*(k$_z$, k$_{z'}$), for LO phonons is then found by:

$$W(k_z, k'_z) = \frac{|D_{op}|^2}{8\pi^2 \rho \omega_0} \int_0^{q_c} \Phi(q_t, q_z) dq_t \, B_\pm(E_{LO} = \hbar \omega_0) \frac{1}{\left| \frac{\partial E(k'_z)}{\partial q_z} \right|_{k'_z = k_p}}, \qquad (3)$$

here *D$_{op}$* is deformation potential of LO phonons for electrons (*D$_{op}$* = 11×10$^8$ eV/cm), ω$_o$ is the maximum phonon frequency of LO phonons or *E$_{LO}$/ℏ*, where ℏ is Planck's constant (ℏ=6.582×10$^{-16}$ eV.s), and *q$_c$* is the maximum allowable value of phonon transversal component (*q$_t$*) within the BZ of bulk silicon which is *q$_c$* = 1.9π/*a* [56]. The last term in equation (3) is density of states evaluated at each available final state (*k$_p$*) into which an electron can scatter. The wave vectors of phonons and electrons are represented by *q$_z$* and *k$_z$*, respectively. These quantities satisfy momentum conservation (See

Supplementary Information and reference [56]). The scattering rates as well as the corresponding final states available after scattering are tabulated according to the index of the secondary sub band as well as the type of scattering, i.e., phonon absorption (ABS) or phonon emission (EM) to be used by EMC code. Figure 2.c shows a simple example of such table.

Within EMC simulation, a uniform electric field along the nanowire axis is applied, and temperature is assumed to be uniform. Initially, electrons are assumed to be distributed uniformly with an equilibrium thermal distribution. As time increase, the evolution of each electron under drift process with multi-phonon scattering is recorded to calculate the electric current in response to the applied voltage (field). At each step of EMC simulation [54], electron transport is confined to the first BZ, which is divided into many $k_z$ grid points (e.g. 8001 points). Corresponding to each point, electron energy, scattering rates and the available final states after scattering, are all recorded and fed into the EMC code from which (a) the time dependent evolution of electron population at each state within the BZ as well as (b) the velocity-electric field plots of the biased SiNWs, are obtained. Details of EMC can be found in reference [57].

The methodology in this work was once used to predict population inversion in strained silicon nanowires and modulation of electron-hole radiative recombination as a result of bandgap and wave function symmetry change [42]. Modulation of recombination rates was later confirmed by experimental works in reference [43]. Later theoretical works on SiGe nanowires also showed the radiative life times of the same order of magnitude as our method [58], which supports the reliability of our approach in predicting the carrier dynamics under the influence of electric field and multi-phonon scattering events.

**RESULTS & DISCUSSIONS**

Figure 2.a illustrates the front view and side view of a 3.1 nm [110] unstrained silicon nanowire terminated by hydrogen atoms. The top and bottom surfaces of the nanowire show canted silicon dihydride (SiH$_2$) units as a result of energy minimization in SIESTA®. The diameter is defined as the average of long and short diameters. The band structure of the unstrained SiNW is shown in Figure 2.d wherein the energy of electrons is plotted against the quasi momentum or wave vector along the periodic length of the nanowire which is $k_z$. The SiNW in this case shows a direct bandgap, i.e., both minima of conduction and valence bands are at the same value of wave vector on BZ center or $\Gamma$ point. Four lowest conduction sub bands numbered from 1 to 4 [Figure 2.d] are those which are used in calculation of electron-phonon scattering under high electric field. This is because the rest of the sub bands are more than $5k_BT \approx 125$ meV above the first sub band minimum and under moderate electric fields they do not play any significant role in our study. The unstrained nanowire has a direct bandgap

of 1.554 eV at BZ center, and there are two indirect sub bands which lie symmetrically at $k_z = \pm 2.67/\pi$ and their energy is 131 meV above the direct sub band minimum. We label this energy difference between minima of direct and indirect conduction sub bands as energy offset $\Delta E$ as indicated in Figure 2.d.

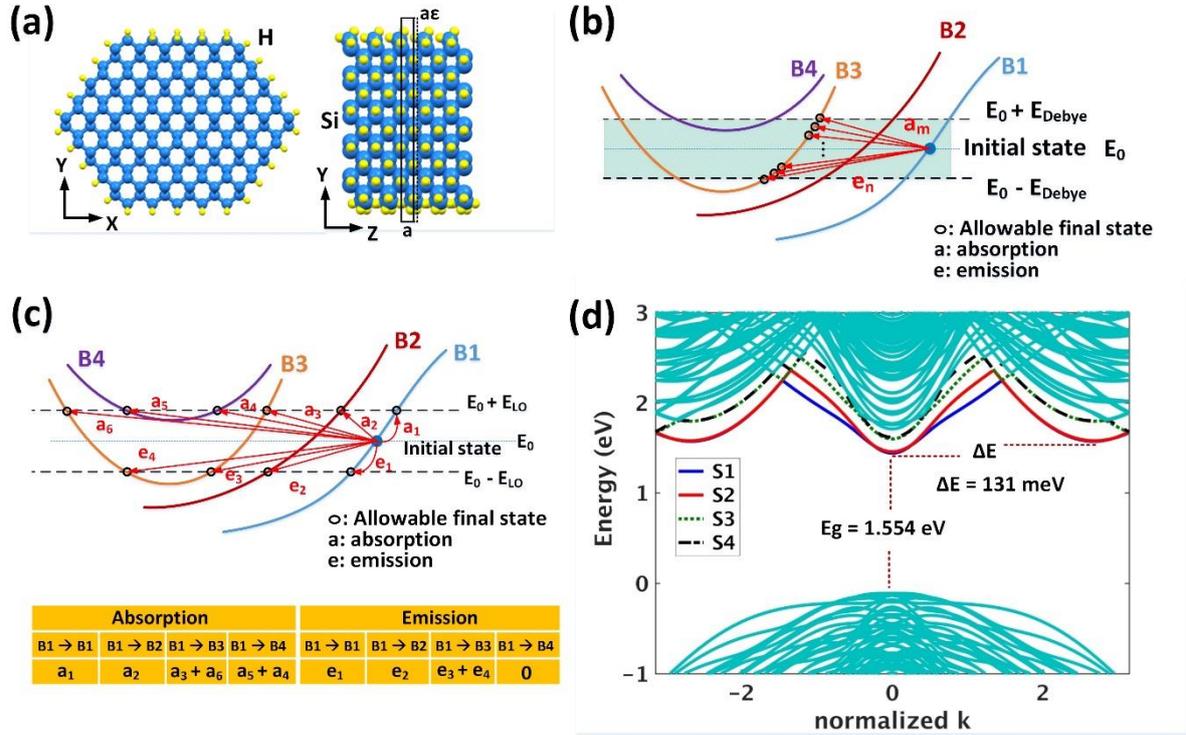

**Figure 2. Atomic structure of an unstrained 3.1 nm [110] SiNW and corresponding electronic (band) structure.** (a) Front view and side view of a 3.1nm [110] SiNW under no strain. Five unit cells of SiNW are shown in the side view highlighting the canted SiH$_2$ (silicon dihydride) units. (b) Graphical representation of finding secondary states for an electron in sub band 1 (B$_1$) which scatters to sub band 3 (B$_3$) assisted by LA phonons. Ideally the secondary states in B$_3$ (shadowed strip) form a quasi-continuum. In simulation, this is limited by the resolution of discretizing the BZ along k$_z$. (c) The same graphical example for scattering from B$_1$ to all sub bands (B$_1$, B$_2$, B$_3$ and B$_4$) assisted by LO phonons. Here there is a limited number of secondary states with corresponding rates arranged in a table. (d) Band structure of the unstrained nanowire showing a direct bandgap value of E$_g$ = 1.554 eV and an energy offset of $\Delta E$ = 131 meV. The first four conduction sub bands (numbered as S1, S2, S3 and S4) are selected to calculate all electron-phonon scattering events required by EMC simulation.

It was previously observed that applying compressive strain lowers the indirect sub band energy and shifts the direct conduction sub band as these bands are made of bonding and anti-bonding *p* orbitals, respectively [42][9, 10]. Therefore, by continuing the application of compressive strain, there is a state at which the direct and indirect sub bands change roles and the bandgap becomes indirect. Hence the conduction sub band minimum and the valence sub band maximum are no longer at the same value of wave vector. Here this transition or sub band flip point occurs at -2 % strain wherein the direct and indirect conduction sub bands are at the same energy level and *ΔE* ≈ 0. After this point, e.g., at -3 % strain (see Figure 3.a), the bandgap is of indirect type and *ΔE* = -19 meV. Figure 3.b and 3.c depict sub bands 1-4 for unstrained and tensile strained (+3 %) SiNWs, respectively. In case of tensile strain, the sub bands respond opposite to the compressive case. Here indirect sub bands experience an upshift in the energy value and direct sub band (at BZ center) experiences a down shift. This results in a higher offset (*ΔE* = 280 meV) than that of unstrained with *ΔE*= 131 meV.

The effect of compressive strain in lowering the energy offset (*ΔE*) may suggest that compressive strain eases the electron transfer from BZ center to the bottom of indirect sub bands with higher effective mass. Hence the onset of negative I-V slope, i.e., a kink in the I-V characteristic could occur at lower values of the electric field. Looking at Figure 3.d suggests otherwise. The plot of velocity (current) versus longitudinal electric field (voltage) shows no sign of kink or negative slope for the SiNWs with small offset, i.e., those at 0 % and -3 %. This is not unexpected by recalling that at room temperature the heavy indirect (high effective mass) sub band must be empty in order to observe NDR. For 0 % and -3 % strained nanowires the ratios of indirect to direct populations are $e^{-\Delta E/k_B T} = e^{+19\,meV/25 meV} = 2.07$, and $e^{-\Delta E/k_B T} = e^{-131\,meV/25 meV} = 0.007$ , respectively. Hence indirect sub bands at these strain values are already thermally filled and mask the occurrence of NDR effect.

On the other hand, Figure 3.d shows that as the nanowires become of more direct bandgap type, it is at +3 % tensile strain that the sought-for NDR emerges. As the effect of applying tensile strain on the band structure is increasing the energy offset (see Figure 3.c with *ΔE* = 280 meV), this suggest the indirect sub band is significantly less populated than those at 0 % and -3 % ($e^{-\Delta E/k_B T} = 4.6 \times 10^{-5}$). After seeing this, now the detailed examination of carrier populations and electron-phonon scattering rates for sub bands is called for.

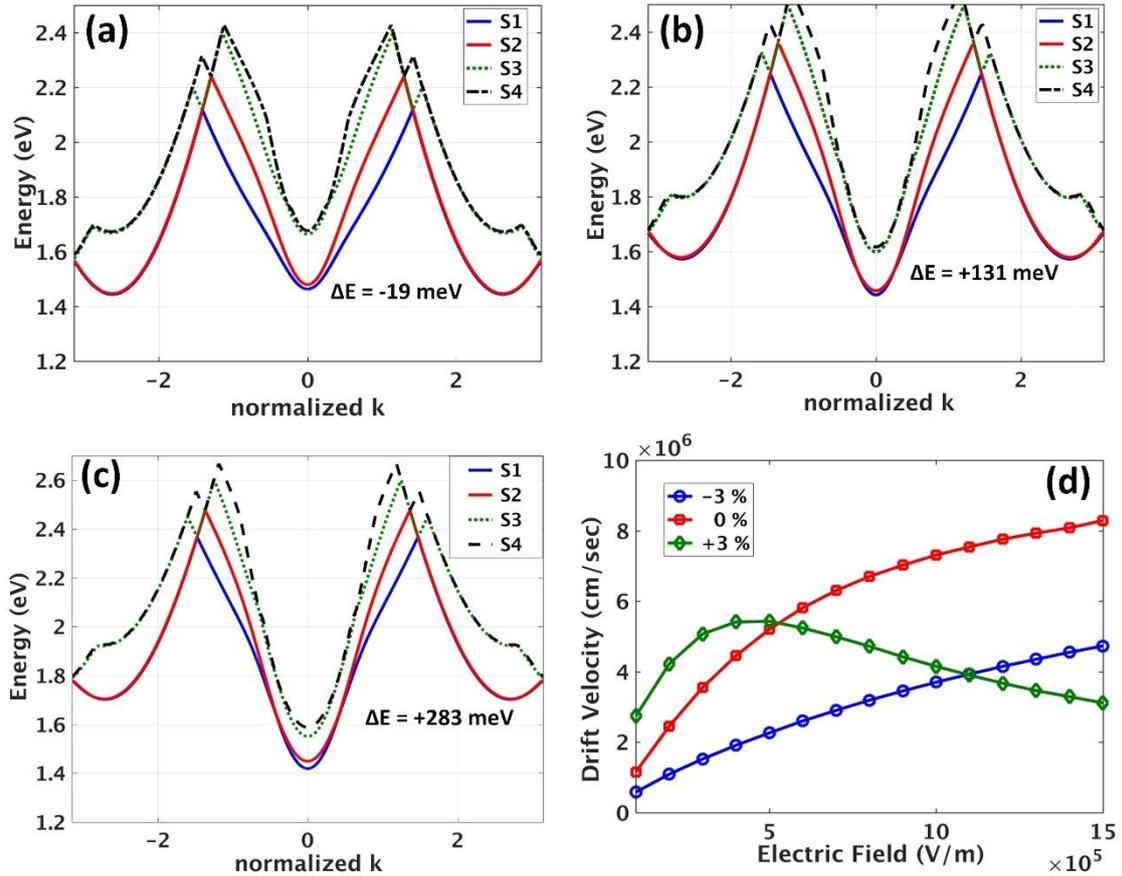

**Figure 3. Conduction sub bands and drift velocity of electrons in response to longitudinally applied electric field under three strain values.** (a) Four conduction sub bands of a -3 % strained 3.1nm [110] SiNW. The energy offset in this case is $\Delta E$=-19 meV, i.e., the bandgap is of indirect type. Panels (b) and (c) show the same sub bands and their new energy values in response to 0 % and +3 % (tensile) strain, respectively. In these cases the offset is $\Delta E$= +131 meV and $\Delta E$= +283 meV, respectively. (d) The drift velocity versus applied longitudinal electric field is a plot proportional to current-voltage characteristics of the device (I-V). Application of tensile strain initiates the current reduction or NDR at $E_{threshold}$ = 5x10$^5$ V/m or 5000 V/cm.

By examining the carrier populations of each individual sub band ($S_1$, $S_2$, $S_3$ and $S_4$ Figure 4) we deduced:

For a -3 % strained nanowire (Figure 4.a), there is no pronounced increase of population in indirect sub bands $S_3$ and $S_4$ (bottom panel) and the populations in sub bands $S_1$ and $S_2$ are changing in unison but with different signs, i.e., as one of them is being filled ($S_1$) the other one ($S_2$) is being depleted. Hence there is carrier transfer from sub band $S_1$ to $S_2$ via all inter-band events for which the scattering rate is around 1 – 10 psec. In other words, both $S_1$ and $S_2$ are within the Debye energy window, facilitating scattering between them rather than scattering (phonon absorption) to $S_3$ and $S_4$, which are at least 200 meV outside the Debye window.

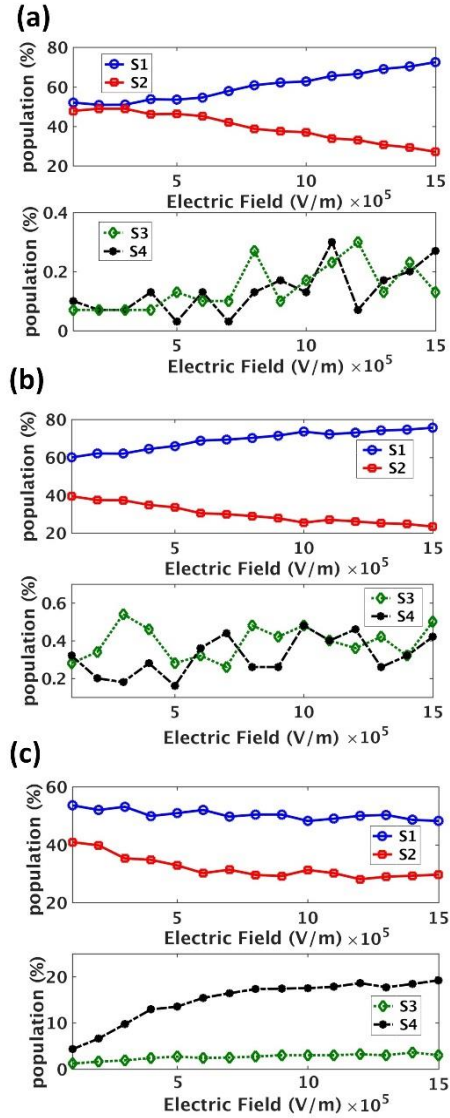

**Figure 4. Evolution of charge carrier (electron) population with electric field.** Sub bands $S_1$, $S_2$, $S_3$ and $S_4$ are shown by blue, red, green and black, respectively. Both in -3 % strained nanowire (a) and unstrained one (b), sub band $S_2$ is being emptied into indirect sub band $S_1$, as the small offset value increases the chance of electron scattering between $S_1$ and $S_2$. Hence there is no significant increase of population in sub bands $S_3$ and $S_4$ which are very high above $S_1$ and $S_2$. For a +3 % strained SiNW (c), both sub bands $S_1$ and $S_2$ are being emptied and high effective mass $S_4$ is populated significantly.

Observing the same behavior as that of Figure 4.a for carrier populations of unstrained nanowire (Figure 4.b), confirms this interpretation. For unstrained SiNW, the indirect sub band $S_1$ and $S_2$ are again being emptied in favor of filling up direct sub band $S_1$ as the distance from indirect ($S_1$ & $S_2$) minima to indirect ($S_3$ & $S_4$) minima is about 200 meV which makes the phonon absorption events to $S_3$ and $S_4$ less probable.

Consequently for -3 % and 0 % strained SiNWs no NDR is observed. However recalling Figure 3.d, and comparing I-V plots for the two cases (blue and red) suggests that indirect bandgap nanowires (i.e. -3

% strained) have higher resistivity than unstrained SiNW by a factor of 2.3. This is because the indirect sub bands of -3 % SiNW are responsible for charge transport and as their effective mass is four times larger than the direct sub band, less mobility or conductivity is expected. Biasing a nanowire with a fixed bias voltage and switching the nanowire between unstrained and compressively strained states will result in a strain sensitive resistor or mechanical force sensor.

Further examination of Figure 3.d reveals that the peak in the velocity-field plot, is shifting to lower electric field values as strain is relaxed from -3 % to 0 %. Finally, at +3 % strain where the offset is maximal, ($\Delta E$ = 280 meV), NDR emerges. This is because after the carriers scatter into indirect sub bands (direct $S_1$ & $S_2$ → indirect $S_1$ & $S_2$), they now have a higher probability to scatter into indirect sub bands $S_3$ and $S_4$ which are now closer as opposed to the rest of sub bands i.e. direct $S_1$ & $S_2$, and direct $S_3$ & $S_4$ (See Figure 4.c). The significant increase in population in indirect $S_3$ and $S_4$ (bottom of panel of Figure 4.c) confirms that *both* $S_1$ and $S_2$ are being emptied (top panel) in favor of transferring carriers to high effective mass indirect $S_3$ and $S_4$ which results in current drop and NDR (see Figure 3.d). Seeing this suggests that increasing the energy offset between direct and indirect sub band leads to reduced backscattering probability for those electrons which travelled to indirect sub bands $S_1$ and $S_2$.

Hence they will find have a higher probability to absorb phonons and end up being in indirect $S_3$ and $S_4$ which are of higher effective mass. Figure 5 shows the evolution of electron population within the BZ of +3 % strained SiNW. It reveals how the population decays from direct sub bands (around BZ center) by increasing the electric field and migrates to indirect ones with higher effective mass which leads to lower velocity.

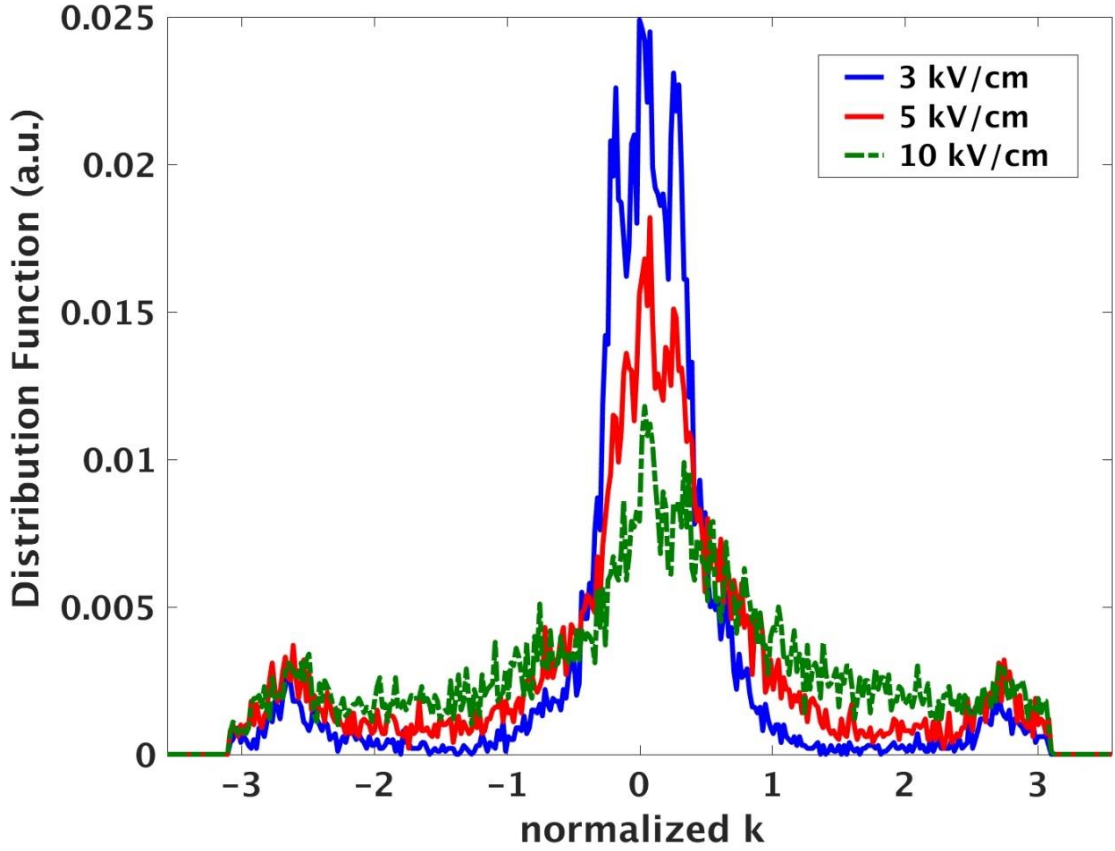

**Figure 5. Evolution of electron population inside the BZ of a +3 % strained SiNW.** Increasing the field results in reduction of the population from BZ center and favors the increase of the same quantity in indirect sub bands and further migration to indirect S$_3$ and S$_4$ with higher effective mass. The overall effect is the current (velocity) reduction leading to NDR.

Referring back to the band structures with small energy offset (e.g. *ΔE* = 131 meV for 0 % strain), it is expected that decreasing the temperature helps to increase the population difference and initiating NDR. For example at *T*= 70K, $e^{-(\Delta E = 131 \text{ meV})/k_B T} = 1.5 \times 10^{-8}$. This implies that small energy offset (*ΔE*) does not preclude the Gunn-Effect provided that the working temperature of the device is lowered. Similar effects were observed for low energy offset InSb and GaSb at *T* =77K [39][59]. We already observed that decreasing the temperature from 300K to 70K drops the direct to indirect electron-phonon scattering rates involving phonon absorption by three orders of magnitude [42]. This unilaterally fills the low energy sub band and enhances the population difference useful for emergence of NDR under high electric fields.

In conclusion there are two main ingredients in order to observe NDR and resulting Gunn Effect in silicon nanowire; (a) A strain dependent band structure in which the sub band energy difference can be controlled by mechanical strain, and (b) Asymmetric electron-phonon scattering channels which work to the benefit of filling up a higher effective mass sub band as the electric field increases.

Moreover, mechanical strain gives freedom to tune or change the onset of NDR initiation or the position of the kink in I-V characteristics with respect to the field. For example embedding silicon nanowires on a deformable substrate, deflecting the substrate can change the band structure. Thus the onset of the NDR or Gunn oscillations will be modulated by strain.

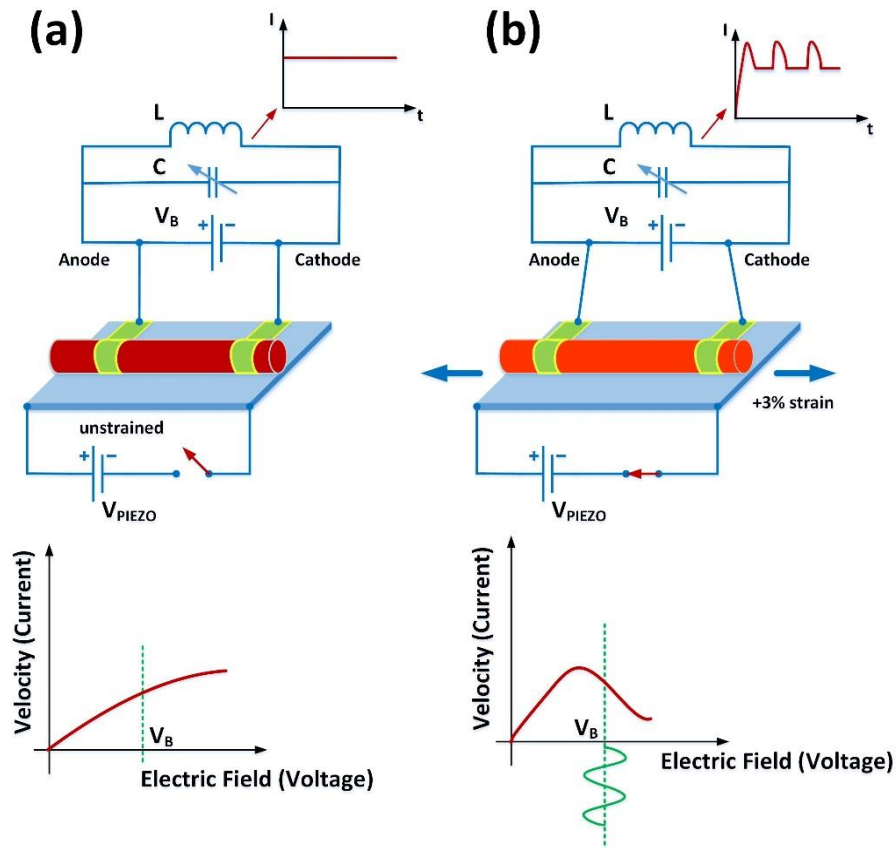

**Figure 6. A proposed experiment where using mechanical strain initiates or inhibits Gunn oscillations in SiNWs.** (a) While the platform is unstrained ($V_{piezo}$ = 0 V) there is no NDR for the applied bias voltage $V_B$. (b) When $V_{piezo} \neq 0$ V, the platform is deflected which strains the wires and $V_B$ finds itself within the range of NDR, hence Gunn oscillations are initiated. The oscillation frequency is adjustable by either strain ($V_{piezo}$), DC bias voltage or electric field ($V_B$) and values of L and C.

Figure 6 sketches such a proposed experiment in which an array of nanowires is mounted on a deformable substrate (e.g. a piezo electrically actuated diaphragm) and biased with a voltage source. In the unstrained state there is no NDR. By deforming the substrate into tensile strain state the onset of oscillations or I-V kink approaches to the desired bias voltage ($V_{threshold}$) and Gunn oscillations will ensue. Reversibly this oscillation can be inhibited by returning to zero or compressive strain regime. The sources of mechanical strain are abundant. Also as we have seen before, even in cases where there is no NDR, the resistivity is modulated by more than 100 % using strain. Another useful experiment could be bending and unbending a long nanowire similar to those fabricated by [43][60], and recording the voltage across the bent sections which are under tensile strain.

Adding spin degree of freedom to this effect will also lead to a rich playground for physical effects and promising spintronic devices [61, 62]. We speculate that in nanowires with large Spin Orbit Interaction (SOI) e.g. InSb, the combination of electric field and mechanical strain will lead to interesting spin-resolved Gunn Effect or oscillations of spin up or spin down currents as they can be initiated under different voltages and strain values (spin resolved Gunn Effect). Alloying SiNWs with germanium is another way of tailoring the band structure and threshold voltage for NDR initiation. The recent techniques of growing SiGe nanowires e.g. in core-shell form are promising [44]. In addition, it was experimentally observed [63] that applying perpendicular magnetic field on bulk InSb samples reduces the Gunn Effect threshold voltage due to Hall Effect. Adding this degree of freedom to the presented theoretical models uncovers new phenomena worth of understanding and consummating to device applications.

In summary using DFT, TB and ensemble Monte Carlo simulations we observed that; **_firstly_**, in contrast to bulk silicon, Gunn Effect is observable in [110] SiNWs under tensile strain of 3 %. This is the result a tunable direct bandgap with high energy offset (*ΔE*) which creates the same mechanism occurring in GaAs, i.e., migration of electrons from low effective mass sub band to indirect high effective mass one as a result of which NDR emerges. **_Secondly_**, the value of threshold field to induce Gun Effect reversible adjustable by mechanical strain. **_Thirdly_**, direct to indirect bandgap conversion in response to the applied strain causes more than 100 % change of resistivity. All the above mentioned observations are promising for more applications of silicon nanowires in the realm of electronic and nano-mechanical devices.

**ACKNOWLEDGEMENTS:**

D.S acknowledges access to the supercomputing facilities provided by Shared Hierarchical Academic Research Computing Network (SHARCNET®) in Ontario, Canada while he was working at University of Waterloo. A.I and D.S acknowledge financial support from Swedish Research Council (Vetenskapsrådet). M.P.A was supported by the National Science Foundation through the Division of Electrical, Communications and Cyber Systems under Grant ECCS-1231927. A.V and R.N acknowledge access to supercomputing facilities of Texas Advanced Computing Center (TACC).